\documentclass[sigconf]{acmart}


\newcommand {\yasmine}[1]{{\color{teal}\bf{}\normalfont}}
\newcommand {\change}[1]{{\textcolor{black}{#1}}}

\usepackage{xcolor}
\usepackage{colortbl}
\definecolor{rowcoloring}{RGB}{183, 226, 240}
\usepackage{rotating}
\usepackage{caption}

\newcommand{\thd}{Tech Help Desk }
\newcommand{\thdns}{Tech Help Desk}

\newcommand{\etal}{\textit{et al.} }
\newcommand{\cf}{Community Forge }
\newcommand{\cfns}{Community Forge} 
\newcommand{\wilk}{Wilkinsburg, PA } 
\newcommand{\bsc}{Business Service Center } 
\newcommand{\bscns}{Business Service Center} 
\newcommand{\cmuns}{Carnegie Mellon} 

\newcommand{\squishlist}{
 \begin{list}{$\bullet$}
  { \setlength{\itemsep}{0pt}
     \setlength{\parsep}{3pt}
     \setlength{\topsep}{3pt}
     \setlength{\partopsep}{0pt}
     \setlength{\leftmargin}{1.5em}
     \setlength{\labelwidth}{1em}
     \setlength{\labelsep}{0.5em} } }
\newcommand{\squishend}{
  \end{list}  }

  \definecolor{lightergrey}{RGB}{225,225,225}


\AtBeginDocument{%
  \providecommand\BibTeX{{%
    \normalfont B\kern-0.5em{\scshape i\kern-0.25em b}\kern-0.8em\TeX}}}

\copyrightyear{2024}
\acmYear{2024}
\setcopyright{rightsretained}
\acmConference[CHI '24]{Proceedings of the CHI Conference on Human Factors in Computing Systems}{May 11--16, 2024}{Honolulu, HI, USA}
\acmBooktitle{Proceedings of the CHI Conference on Human Factors in Computing Systems (CHI '24), May 11--16, 2024, Honolulu, HI, USA}
\acmDOI{10.1145/3613904.3642191}
\acmISBN{979-8-4007-0330-0/24/05}
\begin{document}

\title[Deconstructing the Veneer of Simplicity]{Deconstructing the Veneer of Simplicity: Co-Designing Introductory Generative AI Workshops with Local Entrepreneurs}

\author{Yasmine Kotturi}
\email{ykotturi@cs.cmu.edu}
\affiliation{%
  \institution{Carnegie Mellon University}
  \streetaddress{5000 Forbes Ave}
  \city{Pittsburgh}
  \state{Pennsylvania}
  \country{USA}
  \postcode{15213}
}

\author{Angel Anderson}
\email{angel@forge.community}
\affiliation{%
  \institution{Community Forge}
  \city{Wilkinsburg}
  \state{Pennsylvania}
  \country{USA}
  }

\author{Glenn Ford}
\email{glenn@forge.community}
\affiliation{%
 \institution{Community Forge}
 \city{Wilkinsburg}
 \state{Pennsylvania}
 \country{USA}
 }

\author{Michael Skirpan}
\email{michaelskirpan@andrew.cmu.edu}
\affiliation{%
 \institution{Carnegie Mellon University}
  \city{Pittsburgh}
  \state{Pennsylvania}
  \country{USA}
}

\author{Jeffrey P. Bigham}
\email{jbigham@cs.cmu.edu}
\affiliation{%
  \institution{Carnegie Mellon University}
  \city{Pittsburgh}
  \state{Pennsylvania}
  \country{USA}
}
\renewcommand{\shortauthors}{Kotturi \textit{et al.}}


\begin{abstract}
Generative AI platforms and features are permeating many aspects of work. Entrepreneurs from lean economies in particular are well positioned to outsource tasks to generative AI given limited resources. In this paper, we work to address a growing disparity in use of these technologies by building on a four-year partnership with a local entrepreneurial hub dedicated to equity in tech and entrepreneurship. Together, we co-designed an interactive workshops series aimed to onboard local entrepreneurs to generative AI platforms. Alongside four community-driven and iterative workshops with entrepreneurs across five months, we conducted interviews with 15 local entrepreneurs and community providers. We detail the importance of communal and supportive exposure to generative AI tools for local entrepreneurs, scaffolding actionable use (and supporting non-use), demystifying generative AI technologies by emphasizing entrepreneurial power, while simultaneously deconstructing the veneer of simplicity to address the many operational skills needed for successful application. 
\end{abstract}




\maketitle

\section{Introduction}

\begin{figure}
\centering
\includegraphics[width=5cm]{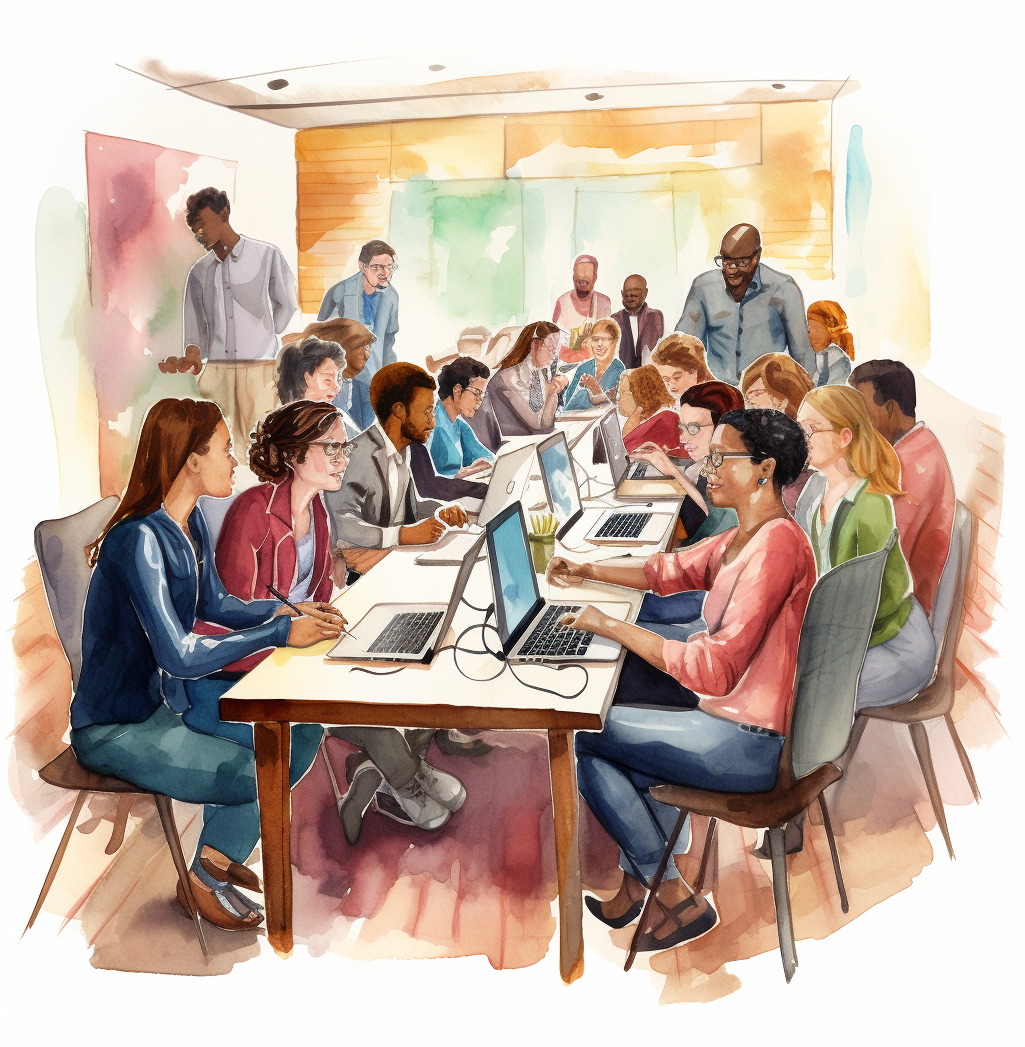}
 \caption{We designed an introductory generative AI workshop series with entrepreneurs and tech providers which centered communal experience, supportive exposure, tangible and actionable exercises, and long-term technical support for maintenance and repair. Image created with Midjourney~\cite{midjourney}.}
\label{fig:label}
\Description{This image, generated with Midjourney, includes a conference table with about 20 entrepreneurs and providers working on laptops and with each other in a room at a conference table. Entrepreneurs and providers appear to be from various age groups, and race and gender identities. This image is in the style of a watercolor painting.}
  \label{fig:1}
\end{figure}\textit{}

The implications of pre-trained, generative AI models--deemed a ``general purpose technology''~\cite{eloundou2023gpts}--for the workforce are vast~\cite{agrawal2019artificial}. 
Local entrepreneurs, who primarily target their local economy and operate at a small scale~\cite{hui2020community}, are uniquely positioned to derive the benefits of these tools by outsourcing tasks, increasing efficiency, and cutting costs~\cite{short2023artificially}. 
\change{For example, consider an entrepreneur with an event planning business and who posts frequently on social media platforms about upcoming occasions, services offered, and prior events; by integrating ChatGPT, Canva text-to-image, and cross-platform management tools, she can quickly generate content and automate publishing workflows, saving her hours worth of work each week~\cite{youtube-canva}.} 
Beyond efficiency gains, equipping local entrepreneurs with generative AI technologies presents a rare opportunity to provide entrepreneurs with support essential to overcome the inertia required to maintain a new venture alone~\cite{mollick2023transformative}.
\change{Whether it's brainstorming with ChatGPT to generate new product ideas or marketing plans, local entrepreneurs can leverage generative AI technologies to overcome creative blocks and catalyze the everyday momentum required for success~\cite{mirowski2023co, swanson2021story}.} 

Yet, despite user-friendly packaging and seemingly simple chat-based interface designs, generative AI technologies like ChatGPT~\cite{bang2023multitask} and DALL-E~\cite{ramesh2022hierarchical} are primarily used by those who have backgrounds in technology or who are college educated~\cite{pew-aug28}. 
Thus, generative AI risks further widening the digital divide~\cite{bbc}, as entrepreneurs who have access to formal and informal education, devices, and technical capital are more quickly learning how to harness this nascent and powerful technology for their benefit~\cite{mollick2023transformative,bell2023entrepreneurship}.
This paper explores how to address this growing disparity by engaging local entrepreneurs in a community setting ~\change{where disparities are heightened given uneven access to technological resources and a history of resource deprivation stemming from post-industrial blight and systemic racial inequalities.}

Previous work in human-computer interaction (HCI) which focuses on local entrepreneurs from lean economies---\change{economies where citizens exude resilience and resourcefulness to overcome minimal resources~\cite{dillahunt2018entrepreneurship}---highlights the critical role of building small, local networks among entrepreneurs when onboarding digital technologies for their business~\cite{dillahunt2018entrepreneurship, hui2018making, avle2019additional, raval2019making, pei2020attenuated, hui2020community, ogbonnaya2019towards, kotturi2022tech}.
These in-person support structures enable entrepreneurs to more effectively vet relationships, reputations, and technological guidance for trustworthiness~\cite{hui2018making,dillahunt2018entrepreneurship}.
Trust is particularly important for entrepreneurs from low-income communities as entrepreneurship is often pursued out of economic necessity~\cite{hui2018making}. 
For instance, small, vetted groups of local entrepreneurs and experts can alleviate the risks and burdens of maintaining technology, and help to foster effective technology use tailored to a specific business domain~\cite{hui2020community}.
In particular, such groups excel when they are informal, non-hierarchical, and showcase all members' expertise~\cite{dillahunt2022village}.
In addition to small groups, one-on-one technology support between entrepreneurs and technology experts can provide needed flexibility to tailor technical advice to the unique backgrounds and domains of local entrepreneurs~\cite{kotturi2022tech}.} 

\change{While prior work details the role of local networks to support maintenance of existing technologies, it is unclear how social support should be structured to onboard local entrepreneurs to generative AI technologies.
In this setting, social support models must respond to fast-paced development and unpredictability~\cite{bommasani2021opportunities}, astounding accuracy alongside hallucinations~\cite{jiang2020can}, and ill-defined notions of AI literacy~\cite{ng2021conceptualizing}, especially in the context of local entrepreneurship.}
Therefore, to inform configurations of social support in this context, we followed a community-driven protocol~\cite{harrington2019deconstructing} building on a four-year partnership with a local entrepreneurial hub dedicated to racial equity in technology and entrepreneurship. 
Together, we co-designed an interactive workshop series across five months to onboard entrepreneurs to generative AI technologies, where workshops were embedded in a monthly entrepreneur mixer~\cite{lu2023organizing} and addressed specific community needs for business support.
To support our iterative design process, we conducted 15 semi-structured interviews with local entrepreneurs and community providers who participated in a workshop(s) and, in doing so, sought to answer the following research questions: \textbf{RQ1} \textit{What is the role of social support when onboarding local entrepreneurs to pre-trained generative AI platforms and features?} Because of entrepreneurs' diverse backgrounds and business domains, as well as their various technological goals and aversions, we asked: \textbf{RQ2} \textit{What are community-driven outcomes of an interactive workshop series intended to onboard local entrepreneurs to generative AI technologies?} By co-designing one configuration of social support which centered community-driven outcomes and values, we then asked: \textbf{RQ3} \textit{How do local entrepreneurs use (and prefer not to use) generative AI technologies for their business? What concerns do local entrepreneurs have when adopting generative AI in their business?}

Three core findings emerged from our analysis. 
First, we found that centering communal experience was critical when onboarding entrepreneurs to generative AI technologies in order to demystify these technologies and mitigate techno-anxieties these tools can elicit when participants' livelihoods are implicated in use. 
Second, while generative AI technologies are often presented with a veneer of simplicity, we detail a laundry list of operational skills beyond prompt engineering that are required for successful use (e.g., browser literacy, successful password management, knowledge of cloud and local storage, keyboard shortcuts).
Through centering within-community expertise, we further detail the steps before and after use of generative AI technologies (i.e., ``pre-'' and ``post-processing'' of inputs to, and outputs of, generative AI technologies) in order for technologies to provide actual and sustained value for local entrepreneurs. 
Third, after introduction, we detail how entrepreneurs used, and preferred not to use, generative AI technologies, which technologies they used, as well as initial concerns entrepreneurs had when using generative AI technologies for their business such as bias and intellectual property infringement.

Taken together, this paper makes the following three contributions. 
First, building on models low-tech social support~\cite{hui2020community,dillahunt2022village}, we present empirical findings for an interactive workshop structure tailored for generative AI technologies that emphasized communal exposure and actionable opportunity for use (and non-use~\cite{baumer2014refusing}). 
In particular, we present an early look at the various operational skills required for AI literacy in the context of local entrepreneurship, and how interactive workshops can support these literacies.
Second, we contribute empirical findings of local entrepreneurs' use of generative AI technologies, as well as their concerns for use as it relates to their business. 
Third, we contribute details of an approach to designing community-driven AI workshops~\cite{cooper2022systematic} that prioritize long-term commitment (e.g., a four year and ongoing tech support program~\cite{kotturi2022tech}), community-driven goals (e.g., workshop series embedded in ongoing community initiatives), and community-centered value generation (e.g., workshop series primarily aimed to support entrepreneurs and improve services within the community center).

\section{Related Work}
Three bodies of scholarship motivate our work: (1) entrepreneurialism in the digital age, and the importance of social support when onboarding local entrepreneurs to digital technologies, (2) generative artificial intelligence, and recent HCI research on how people use generative AI technologies (e.g., large language models), and (3) community-driven research methodologies in computing research.
 
\subsection{Entrepreneurialism in the Digital Age}
Entrepreneurship in the 21st century takes various shapes from tech entrepreneurship~\cite{bailetti2012technology} to side-hustles managed entirely on mobile devices~\cite{pew_research_demographics_2019}.
Digital platforms and tools enable a wide range of entrepreneurial pursuits which are less dependent on brick and mortar storefronts and local material sourcing and manufacturing~\cite{agrawal2013digitizationoflabor}.
Yet, while entrepreneurialism in the digital age is touted to be a democratizing force~\cite{von2006democratizing}, scholars increasingly critique the burden placed on entrepreneurs to keep pace with technological innovation~\cite{irani2019chasing}.
In particular, pressures of digitization disproportionately affect entrepreneurs in resource-constrained communities or ``lean economies'' due to lack of access to technology resources, education, and capital~\cite{hui2018making}.
These effects are further compounded based on race and class due long-standing systemic and institutionalized racism and classism in the U.S.~\cite{benjamin2023race}.
To better understand these inequities, prior work has detailed the types of technology challenges faced by local entrepreneurs in lean economies~\cite{dillahunt2018entrepreneurship,hui2018making,raval2019making,pei2020attenuated,hui2020community,ogbonnaya2019towards,avle2019additional}---entrepreneurs who primarily engage their local economy to overcome a lack of job opportunity and upward mobility~\cite{hui2018making,hui2020community}---who are often driven by economic necessity rather than choice~\cite{hui2018making}; such challenges include a lack of trust in technology platforms due to tech-based erasure and harm~\cite{dillahunt2018entrepreneurship}, unsupported technical skill acquisition alongside constant need for ``upskilling''~\cite{kotturi2022tech}, unreliable devices and maintenance difficulties~\cite{hui2020community}, and more. 

\subsubsection{The Importance of Low-Tech Social Support Among Local Entrepreneurs}
To overcome barriers to economic mobility, HCI scholars have detailed the importance of low-tech social support among local entrepreneurs. 
For instance, Dillahunt~\etal presented a model of social support for individuals experiencing financial hardships called ``the Village''---a community-based mentorship model which centered non-hierarchical relationships in non-institutional settings~\cite{dillahunt2022village}. 
In doing so, they differentiate a village model of mentorship from predominant mentorship models that exist primarily in the workplace and educational settings, and that assume expert-novice relationships (such as in the case of legitimate peripheral participation~\cite{lave1991situated}). 
In the context of poverty-stricken adults in the United States, the authors found that in-person interactions and trust building were required to facilitate economic mobility, and that technological mediation of relationships may prohibit such relational foundations~\cite{dillahunt2022village}.
In the context of local entrepreneurship, entrepreneurs in resource-constrained communities similarly preferred to become digitally engaged by leveraging social networks of peers and experts~\cite{hui2018making,hui2020community}. 
Hui~\etal articulated one model of social support called ``Community Collectives'' which involved small informal and formal groups of like-minded individuals who joined together in a collective pursuit to become entrepreneurs, specifically local tour guides~\cite{hui2020community}.
In their analysis, they found that digital platforms assumed access to basic resources, and that low-tech social supports were critical (e.g., resource-connecting organizations, regular in-person meetings, and paper planning tools). 

While the ``Community Collectives'' model focused on entrepre-neurs with a shared business domain (i.e. local tourism),  ``Tech Help Desk'' provided a strategic and relational model of technical support for entrepreneurs from diverse backgrounds, business domains, and tech preferences through one-to-one, in-person, and long-term technical support ~\cite{kotturi2022tech}.
Technical support staff worked collaboratively with entrepreneurs to solve the ``long tail of computing challenges'', or a large number of distinct challenges that were surfaced and solved. 
In doing so, this work detailed the various digital literacies entrepreneurs needed to become digitally engaged which included both strategic skills---high-level goal setting---and operational skills---low-level implementation~\cite{van2011internet,ala2011mapping}. 
This paper builds on this scholarship in two ways; first, by investigating the role of social support when onboarding local entrepreneurs to generative AI platforms and features. 
Second, this paper then investigates which operational and strategic skills may be helpful for local entrepreneurs in order to apply these technologies to their business pursuits. 

\subsection{Generative Artificial Intelligence}
Generative artificial intelligence has introduced a paradigm shift in computing, illustrated through the recent deployments and high adoption rates of pre-trained models wrapped in user-friendly platforms such as Midjourney~\cite{midjourney} and ChatGPT~\cite{bang2023multitask}. 
Large language models (LLM) like GPT exhibit high task performance with minimal training~\cite{bang2023multitask, ramesh2021zero}. 
Recent interface designs which support end-user interactions with large language models, such as ChatGPT, rely on prompt-based interaction techniques where a user providers a set of instructions which can be written with natural language or code-like syntax~\cite{liu2023pre}.
In addition to text generation, text-to-image generation is yet another paradigm shifting technology deployed for public consumption within the last year, such as diffusion models~\cite{ramesh2022hierarchical}.
Platforms like Midjourney and DALL-E 2 leverage such models to enable high-fidelity imagery generation by end-users~\cite{midjourney,ramesh2022hierarchical}.

\subsubsection{Prompt Engineering}
A critical part of effective use of generative AI technologies is ``prompt engineering'' or writing natural language instructions that models respond to~\cite{bach2022promptsource}.
Researchers have detailed prompting techniques such as personas (i.e., a user provides a LLM with persona or role to play when generating output), flipped interaction (i.e., a user requires a LLM to ask questions rather than generate output), context manager (i.e., a user specifies the context for a LLM's output)~\cite{white2023prompt}, meta-prompting (i.e., a user asks a LLM to create its own prompt~\cite{reynolds2021prompt}), repetition in prompts~\cite{ramesh2021zero}, giving examples of desired interaction~\cite{brown2020language,openai}, adopting code-like syntax and structure~\cite{bach2022promptsource}, and even adopting a well-known Q\&A structure from online forums called ``ask me anything''~\cite{arora2022ask}.

\subsubsection{HCI Applications of Generative AI}
HCI researchers are rapidly charting the design space of generative artificial intelligence~\cite{morris2023design}, as well as developing applications of generative AI. 
For instance, recent work designed novel interaction modalities~\cite{park2023generative,park2022social} and prompt chaining (where the output of one prompt automatically becomes the input in another prompt~\cite{wu2022ai}), as well as applying large language models to support idea generation in the context of creative writing~\cite{gero2022sparks,swanson2021story} and script writing~\cite{mirowski2023co}.
In addition, HCI researchers have studied effective prompt engineering such as in the context of software engineering~\cite{jiang2021genline, zamfirescu2023johnny}. 
For instance, Zamfirescu-Pereira~\etal studied how non-AI experts ``intuitively'' approached chat-based LLMs and constructed prompts for programming tasks~\cite{zamfirescu2023johnny}.
They found that end-users approached prompt designs opportunistically and were overconfident, without a clear strategy nor assessment protocol.
Therefore, the authors called for further studies with users from more diverse backgrounds, noting that even though their users were non-AI experts, they were, in fact, graduate students or professionals in STEM-related fields. 
The authors also suggested an area for future work: how can tools help set expectations for end-users to make them more accurate? 
Such inquiries are relevant to this paper, as existing reference guides---even those dedicated to ``beginners''---require a high level of technical knowledge to parse and make actionable~\cite{openaiplayground,openaiexamples}.

While there is limited scholarly work specifically applying generative AI in the context of entrepreneurship, there is a rapidly growing body of tangentially-related scholarship which may be applicable for local entrepreneurs, such as using generative AI to increase productivity~\cite{noy2023experimental} and improved creative outcomes with ideation support~\cite{haase2023artificial}, such as personalized brand material~\cite{wang2023popblends}.
Alongside these empirical results, there are hundreds of accounts and videos of self-proclaimed entrepreneurs and influencers providing advice online for \textit{how to use AI for your business}, often providing demos of ChatGPT~\cite{youtube-chatgpt}, DALL-E 2~\cite{youtube-dalle}, Canva's Text-to-Image feature~\cite{youtube-canva}, Vidyo.ai~\cite{youtube-vidyo}.
Yet, despite user-friendly packaging and seemingly simple chat-based interface designs, generative AI tools like ChatGPT and DALL-E are primarily used by those who have backgrounds in technology or who are college educated~\cite{pew-aug28}.
Thus, there is a further widening of the digital divide~\cite{bbc}, as entrepreneurs who have access to formal and informal education, devices, and technical capital are rapidly learning how to fully harness this nascent and powerful technology for their benefit~\cite{mollick2023transformative,bell2023entrepreneurship}.
This paper considers how to address this growing disparity by investigating the skills that are presumed to be implicit knowledge among end-users.
In other words, this paper aims to detail the ``long tail'' of technical skills~\cite{kotturi2022tech} that are required to use generative AI technologies effectively but are often overlooked.

\subsubsection{Dangers, Ethics, and Responsibility}
The opportunities presented by the emergence of generative AI technologies come with many ethical concerns and potential downfalls as we careen into a new paradigm of human-computer interaction, or ``human-centered AI''~\cite{hai}. 
Even in the short period of time these technologies have been available, manifest risks have emerged~\cite{bommasani2021opportunities}.
Scholars have detailed issues such as nonfactual responses and misinformation~\cite{jiang2020can}, toxicity~\cite{gehman2020realtoxicityprompts}, ethical, legal and environmental concerns~\cite{bender2021dangers}, and more.
Such issues stem in part from model training techniques which often lack human oversight due to the large scale.
For example, since the corpus of text used to train LLMs is riddled with the same biases found across the internet, stereotypes and prejudice sentiments are inculcated into models that repeat and reinforce these harmful viewpoints~\cite{bender2021dangers}.

With many risks present and looming, questions may be posed on the responsibility and ethics of propagating these technologies into new spaces, especially when the digital and AI literacy in a space may be inadequate for minimizing these risks.
However, the reality of the digital divide is that we diminish human autonomy and thus human flourishing when perpetuating an access gap.
Our stance is that we need to take responsibility and be honest about the risks and limitations of technology when sharing with new people.
We also believe that we must take an approach to this work that is empowering for new users, and focus on increasing the agency of people who are otherwise more likely to miss out on technological advancements.
In doing so, we situate our work in the larger discourse of responsible AI~\cite{lee2020human}, specifically by considering how to introduce generative AI technologies into new spaces in ways that foster critical conversations and community empowerment to use (and not use) these novel technologies.


\subsection{Community-Driven Research in Computing}
When engaging underserved communities in the design of computing technologies, standard user-centered methods often falter as they assume a positive relationship between researchers and participants~\cite{dillahunt2017reflections}, encode infantilizing treatment of participants~\cite{harrington2019deconstructing}, and may perpetuate forms of institutional racism~\cite{tran2019gets,williams2019co}.
Community-based research, where community members and researchers work in tandem to conduct research and derive solutions, can result in outcomes which center community ideas, assets, desires, and needs~\cite{wong2022elevating,israel1998review}.
Doing so successfully requires an awareness of the power dynamics at play across stakeholders~\cite{sabiescu2014emerging}.
This can also require reorienting the traditional HCI paradigm, and providing support for non-experts to actively shape research objectives~\cite{winschiers2013toward}.
When community stakeholders assume a more directive role in research processes, community-driven collaboration can hearken community wisdom and showcase alternative types of knowledge not traditionally surfaced in the design process~\cite{cooper2022systematic,le2015strangers,tandon2022hostile}.
In particular, Lu~\etal detailed how community events can play a critical role in fostering participatory action research with underserved communities~\cite{lu2023organizing}.
The authors leveraged community-driven events to spark conversation about community surveillance in order to center lived experience~\cite{lu2023shifting} and bolster participatory noticing through photovoice~\cite{lu2023participatory}.
By designing for the context of community events, the authors prioritized meeting community stakeholders where they are, both physically in local community centers and on a topic which was pertinent to the community. 
Therefore, our work draws on community-driven scholarship to consider the role of community events when fostering ongoing conversations around generative AI technologies and their implications in the context of entrepreneurship.

\section{Methods}

\subsection{Location and Site}
We conducted our research within a coworking space and community hub for entrepreneurs based in \wilk called \cfns. \\

\noindent \textbf{\wilk} Wilkinsburg is a borough of Allegheny County. 
The population of Wilkinsburg is roughly 49.5\% Black and 23\% of people are living at or below the poverty line~\cite{wilk}.
Wilkinsburg immediately borders but is not part of Pittsburgh, and it is one of the many unincorporated municipalities that acutely struggles with resource deprivation and long-term disinvestment~\cite{allegheny}.
In 2021, the Pittsburgh Metropolitan Area was considered to be one of the U.S.'s ``Apartheid Cities''~\cite{pgh}, as the structures of power within the city continue to perpetuate systemic racial inequality and injustice~\cite{massey1993american}, magnified by the post-industrial blight the region experiences. \\ %

\noindent \textbf{\cf} \cf is a former elementary school repurposed into a space that hosts mixed programming geared towards developing a more equitable economy for Wilkinsburg and the Greater Pittsburgh Metropolitan Area.
Towards this goal, \cf provides financial resources, jobs, job training, business development, youth empowerment programs (e.g., courses, summer camps, hands-on-learning), and community outreach events (e.g., food and supply giveaways, music and movie nights), and voting resources.
\cfns's business development resources include: coaching and professional service referrals, technical assistance, networking opportunities, financial support, and affordable office rentals (repurposed classrooms with coworking and individual office space). 
\cf works with roughly 50 local businesses each year through a variety of programs where 95\% of the businesses are Black-owned, approximately 90\% of entrepreneurs do not have a college degree, and 80\% are first-time entrepreneurs. 
To spread information about resources available within the space, \cf relies on word-of-mouth and social media, as well as working with existing organizations in Wilkinsburg and Pittsburgh which support entrepreneurs. 
\cf also hosts monthly entrepreneur nights, where local entrepreneurs can network, enjoy free food, share updates and hear any announcements with the space.
\change{To support itself and provide various programming for the community, 65\% of \cfns's budget is earned revenue from a mixture of sources such as building revenues (i.e., leases and rentals of space for coworking), government contracts, school contracts, and other partnership contracts.
The remaining 35\% of \cfns's budget is funding from philanthropic foundations.}\\

\noindent \textbf{\bsc and \thd}
\cfns's resource center for entrepreneurs, called the \bscns, provides a variety of services at a subsidized rate including accounting, bookkeeping, and marketing services. 
In addition, to provide technical support for entrepreneurs and residents, the research team and \cf leadership collaboratively designed \thdns, which has been running for four years and is free for entrepreneurs to use~\cite{kotturi2022tech}. 
\thd connects local engineering Ph.D. students, trained in community-based methods, with entrepreneurs to provide weekly technical support for a range of computing issues such as website building and design, file organization and management, cybersecurity monitoring, and more. 
To date, the service has provided technical support to over 70 entrepreneurs addressing over 200 distinct computing issues. 
Central to the success of \thd is the emphasis on relationship building and trust throughout long-term and reliable technical support. 
As described in the following sections, these ongoing services provided a steady foundation that the academic-community partnership relied on for successful implementation of the workshop series.

\subsection{Co-Designing Introductory Workshops to Generative AI for Local Entrepreneurs}
To facilitate the co-design process of the generative AI workshop series, the academic team members and community stakeholders (i.e., staff and leadership at \cfns) met weekly April 2023-September 2023. 
Early meetings outlined community goals, while later meetings served as a way to refine goals and continue iteration of workshop design based on entrepreneurs' and providers' feedback. 
At a high-level, \cf leadership wanted workshops to be a multi-part series, where each workshop focused on a separate, yet connected topic. 
Because of the prevalent need for business marketing assistance for entrepreneurs participating in the \bsc and \thdns, workshops focused on using generative AI for marketing and branding \change{(See Table~\ref{tab:workshopoverview}).}

\subsubsection{Workshops Goals}
\label{sec:workshopgoals}
\cf leadership and research-ers co-articulated three primary goals of the workshop series: first, the workshop series needed meet entrepreneurs where they are in terms of their level of comfort and trust with technology---or lack thereof---and provide support beyond technological means (e.g., provide food for workshops during dinner time). 
It was especially important to frame engagement with technology as highly optional to support non-use for those who were uninterested~\cite{baumer2014refusing}, and facilitate other activities alongside such as peer networking (See Table~\ref{tab:workshopoverview}). 
Second, the workshop series needed to be actionable and tangible: entrepreneurs needed to be able to do hands-on work (rather than solely listening to lecture-style presentations), and the work needed to be directly tailored to their business (as opposed to generic assignments). 
Finally, the workshop series needed to be embedded in a network of trust. 
In this way, the workshop series needed to be hosted at \cf (rather than at \cmuns), and draw on existing technical services already present within \cf which have built a reputation for providing trustworthy technical support. 
By embedding the workshop series in other programming within \cfns, this also meant the workshops better connected entrepreneurs to ongoing support for them to access between and after workshops.
\change{Another way we prioritized trust building was by supporting various levels of engagement in workshop and research activities.
In particular, as done in prior work with service and event-based community engagement~\cite{lu2023organizing,kotturi2022tech}, participation in the study was optional, and attendees could opt-in to the study after meeting with the research team, asking questions about the research process, compensation, and so on.
}

The first two workshops served as a soft launch for the latter two workshops and were more informal as a way to gauge entrepreneurs' initial reactions, interactions and preferences when it came to using generative AI for their businesses.
To facilitate effective iteration of workshop structure, the \bsc conducted pre- and post-questionnaires with attendees. 
Pre-questionnaires asked entrepreneurs to share their level of experience with various generative AI tools. 
Post-questionnaires solicited feedback from entrepreneurs as to what were the most and least valuable aspects of the workshop, among other kinds of internal data collection used to improve the \bsc and \thd services in concert with the generative AI workshops. 
In addition, providers were asked to complete a feedback form after workshop completion to provide quick feedback on the event. 
Part of evolution focused on fine-tuning the particulars of the workshop format. 
For instance, initial workshops provided entrepreneurs with prompt libraries, both paper and digital copies~\cite{hui2020community}, as exploring prompt libraries is often a recommended approach for initial use~\cite{openaiexamples}.
But \cf leadership and community providers noted how these libraries, while applicable and tangible, presumed several critical steps of knowing how to situate prompts in the context of use, indicated by entrepreneurs' overwhelm when presented with prompt libraries. 
Therefore, latter workshops shifted away from prompt libraries upon introduction, and instead relied on community providers to facilitate co-articulation of prompts with entrepreneurs upon discussion with entrepreneurs about their business and tech goals.    

Over the course of the workshop series, training materials were assembled and distributed to providers as an overview of learning objectives (e.g., prompt engineering techniques and examples such as personas, flipped interaction, repetition~\cite{white2023prompt}).
Learning objectives were meant to provide guidance when needed, but providers were encouraged to customize their approach to working with each entrepreneur in order to meet entrepreneurs' unique needs.  
Examples of learning objectives included breadth-oriented learning objectives such as ``Entrepreneurs are aware of at least three AI tools or features they can use to create images for this business (e.g., Canva, Pixlr, DALL-E 2, Midjourney),'' and depth-oriented objectives such as ``Entrepreneurs know how to write text-to-image prompts which create useful images for their business branding.'' 
All learning objectives were paired with a measurable outcome. 
In the final workshop, there was a live demonstration by a local entrepreneur who used generative AI for his apparel business; specifically, he used AI image-generation (DALL-E 2) to create custom designs.


\begin{table*}[h]
\centering
\small
{\color{black} 
\begin{tabular}{ |p{2.5cm}|p{2.7cm}|p{2.7cm}|p{2.7cm}|p{3cm}| } 
 \hline
 \rowcolor{rowcoloring}
  & \textbf{Workshop \#1} & \textbf{Workshop} \#2 & \textbf{Workshop \#3} & \textbf{Workshop \#4} \\ 
 \hline 
 \textbf{Workshop date and duration} & May 2023, 2 hours & June 2023, 2 hours & July 2023, 2 hours & August 2023, 2 hours \\
 \hline
 \textbf{Workshop theme} & Marketing (SEO)  & \raggedright Marketing (Social Media) & \raggedright Marketing (Copy and Email) & Marketing (Images and Branding) \\
 \hline
 \raggedright \textbf{Number of attendees} & \raggedright 9 Entrepreneurs \& 4 Providers & \raggedright 7 Entrepreneurs \& 3 Providers &  \raggedright 10 Entrepreneurs \& 4 Providers  &  10 Entrepreneurs \& 7 Providers \\ 
 \hline
\raggedright \textbf{Generative AI Technologies covered} & ChatGPT, Bard & ChatGPT, Bard & ChatGPT, Bard & DALL-E 2, Canva Text-to-Image App, Pixlr, Midjourney \\
 \hline
 \raggedright \textbf{Additional activities offered} & Board games, peer networking & Record business pitch, peer networking & \raggedright Informal tours of \cfns, peer networking & In-house demonstration of DALL-E 2, peer networking \\
 \hline
\end{tabular}
}
\caption{\change{Overview of the workshops series. Given the need for marketing support among entrepreneurs at \cfns, all workshops focused on different aspects of small business marketing. However, providers were encouraged to work on whichever tasks were most pertinent to the entrepreneurs and the low provider-entrepreneur ratio supported this flexibility. The additional activities offered were essential to support non-use, and food from minority-owned restaurants was provided.}}
\label{tab:workshopoverview}
\end{table*}

\subsection{Workshop Series Overview}
\label{sec:workshopdesign} 
In total, we offered four workshops on the third Wednesday of May, June, July and August of 2023 in the evening, 5:00-7:00 PM \change{(See Table~\ref{tab:workshopoverview}).} 
All workshops were embedded in \cfns's monthly ``Entrepreneur Night'' which had been running since March 2022.
Workshop attendance was typically 50\% of event sign-up rate; providers were recruited based on sign ups to achieve a 2:1 ratio or lower between entrepreneurs and providers.
Workshops included 30 minutes of meet-and-greet time with other entrepreneurs and providers. 
Next, entrepreneurs and providers gathered and everyone briefly introduced themselves and their business, as well as responded to an ``AI Icebreaker'' prompt such as: ``share an emotion that arises when you think of using AI for your business''. 
Then, entrepreneurs and providers formed small groups to begin a one-hour interactive portion of the workshop, co-articulating and iterating on prompts relevant to the workshop's theme. 
The layout of this interactive portion of the workshops evolved from having small groups of providers and entrepreneurs distributed across a large room to sitting side-by-side at a large conference table in order to better support sharing (See Figure ~\ref{fig:1}).

Each workshop included devices which were ready for use: iPads and Dell Laptops. We set up free accounts and created paid accounts when needed, associated with \cfns, and signed in to all relevant platforms on each device. 
Entrepreneurs typically had paper and pen note-taking tools with them, or they were provided with this if not.
At the end of the workshop, entrepreneurs were encouraged to share what they had created during the workshop with the group. 
See supplemental materials for an event timeline example.
\change{Workshops were not recorded and our analysis of the workshops was based on the interview participants' reflective experience after their participation in a workshop(s), as described in the next section.} 

\subsection{Interviews with Local Entrepreneurs and Community Providers}
We conducted 15 interviews (ranging from 30 minutes to two hours long) with seven local entrepreneurs and eight community providers who participated in one or more of the workshops.
We interviewed all but two community providers who participated in the workshop series.
\change{We recruited entrepreneurs from the workshops by announcing the opportunity to everyone during the workshops and posting in the \cf entrepreneur Facebook group; we did observe that entrepreneurs who responded to interview requests were typically those who were highly engaged during the workshops (e.g., they attended multiple workshops, asked the most questions during the workshop, or came to \thd between workshops).}
Participants were compensated \$20/hr. 
Interviewing both entrepreneurs and providers helped to gain a more well-rounded understanding of the workshops. 
See supplemental materials for full interview protocols.  

\subsubsection{Participants}
On average, the seven entrepreneurs we interviewed participated in 1.4 workshops. 
These entrepreneurs had various product and service-based companies, with little overlap in domains such \change{as a podcast producer, gift basket maker, event planner, clothing designer, candle maker, and more} (See Table \ref{tab:demo}). 
Participants' demographics reflected the communities that \cf aims to support: individuals with low to moderate income, who primarily engage in entrepreneurship out of necessity or to overcome a lack of local job opportunities. 
Participants' age ranged from 22 of over 65. 
Community providers included \cf staff from the \bscns, \thdns, as well as the youth tech programs (See Table \ref{tab:demo}). 
Providers had post-graduate degrees in education, business or computer science. 
Academic volunteers had training in community-based research methods and had been vetted by community stakeholders. 

\subsection{Data analysis}
For semi-structured interviews, the research team conducted audio recording and took detailed field notes. 
13 out of 15 interviews were conducted remotely via Zoom.
All audio recordings were transcribed (with Zoom or Temi transcription services).
We followed participant quote editing conventions consistent with applied social science research practices~\cite{corden2006using}---i.e., removed filler words and false starts, and re-punctuated and used ellipses to indicate substantial omissions. The research team analyzed these data through a process of open coding to identify initial themes across the interviews~\cite{charmaz2007grounded}.
The first author wrote analytic memos for each interview with an average word count of 810 words~\cite{charmaz2007grounded}, which summarized the interviews along three themes: feedback on workshop structure and experience, skills needed to use generative AI for business application, and uses of AI (and concerns). 
All memos were reviewed by our community partner, serving as a member check~\cite{charmaz2007grounded}.

\subsection{Community-Driven Research Process}
In community-collaborative approaches, it is critical that community partners are involved in all stages of research processes~\cite{cooper2022systematic}.
We centered community directives in the following ways: university and community teams had a four-year working relationship established through running \thdns.
By having research team members physically at \cf every week for four years, this created effective work relations and clear understanding of intent. 
\change{In addition, the IRB protocol was informed by the four year relationship between the university team and \cfns, where the university-community team co-articulated the protocol's structure as done in prior work~\cite{boser2006ethics}.
In particular, participation in the study was not required in order to access technical services, and data collection was community-driven and connected to ongoing programming at \cfns.
Community partners received ethics certifications, which was paid for by the university team.
}
Lastly, the first two authors---each a representative from the university team and community partner---worked together throughout each stage of the collaboration to design workshops, interview protocol, conduct and analyze interviews together, and engage in reflexive discussion. 

\subsection{Positionality Statement}
We disclose the identities and positionality of the researchers and authors of this paper, as a concern for reflexive design research practice~\cite{liang2021embracing, schlesinger2017intersectional}.
This research team comprised one white woman (a U.S. immigrant from Canada); two white men from the rural Midwest of the U.S. and an impoverished, post-industrial part of Eastern U.S.; one person who identifies as a triple minority as a non-binary, queer person of color from the Eastern U.S; one African-American man who is neurodivergent and is from a low-income background.
The research team comprises two researchers who are upper management at the field site, two staff members at the field site, and three researchers in a technical department at a private U.S. university.
In particular, we note how the three middle-aged, white researchers do not have certain lived experiences that are relevant to this study such as the impact of forms of violence due to racism, ageism, or xenophobia (especially in the context of technology education). Given the predominantly white research team, we took measures to mitigate power imbalances and to cultivate a more equitable relationship between the research team and \cf members (as well as within the research team). 
For instance, all members of the research team committed to the \cf mission statement (exhibiting such commitment through a four year working relationship through hosting \thdns), prioritized generating immediate value for the community members rather than optimizing the research agenda, maintained transparency with research practices, deprioritized data collection, and routinely sought feedback from \cf members and staff.

\begin{table*}
\small
\Description[Participant demographics]{This table provides an overview of participants, their job or business description or business type,  duration of using AI/ML technologies, as well as which tools they use and for which business purposes.}
\caption{Overview of participants, their job or business description (community provider ``P\_'') or business type (local entrepreneurs ``E\_''),  duration of using AI/ML technologies, as well as which tools they use and for which business purposes. Entrepreneurs highlighted in blue. Note that E3 provided the expert demonstration described in Sections \ref{sec:workshopgoals} and \ref{sec:inhouse2}.}
\begin{tabular}{|l|p{2.45cm}|p{3cm}|p{3cm}|p{3cm}|p{3cm}|}
\toprule
\textbf{ID} & \textbf{Participant Type} & \raggedright \textbf{Job/Business Description} & \raggedright \textbf{Prior experience with AI} & \raggedright \textbf{Uses of Generative AI} & \textbf{Main Tools Used} \\ 
\midrule
P1 & \raggedright Community Provider & \raggedright Educator, \cf leadership & 1 year & Portraits, media, branding materials & Wonder AI, ChatGPT \\
P2 & Community Provider & \raggedright Educator, \cf staff & 3 months & Flyers, media & Wonder AI, PhotoRoom \\
\rowcolor{rowcoloring}
E3 & Local Entrepreneur* & Clothing Brand, \cf fellow & 1 year & \raggedright Podcast editing and imagery for products & DALL-E 2, ChatGPT, Nomad Sculpt \\
P4 & Community Provider & Maintenance Technician, \cf staff & 3 months & Video and audio file editing & Capcut, Vidyo.ai \\
\rowcolor{rowcoloring}
E5 & Local Entrepreneur & Youth mentorship & 3 months & \raggedright Website copy, editing video/audio content, hashtags & ChatGPT, Vidyo.ai \\
\rowcolor{rowcoloring}
E6 & Local Entrepreneur & Luxury candles & None & Grant proposal, SEO, lead generation & ChatGPT, DALL-E 2 \\
P7 & Community Provider & \raggedright Educator, \cf leadership & >5 years & Research, education & ChatGPT \\
\rowcolor{rowcoloring}
E8 & Local Entrepreneur & \raggedright Notary and podcast producer & None & Podcast scripts, logos, flyers & ChatGPT, DALL-E 2 \\
\rowcolor{rowcoloring}
E9 & Local Entrepreneur & Gift basket and party planning & None & \raggedright Website copy, flyers &  Wix Copy Generator, Canva Text-to-Image, Copy.ai \\
\rowcolor{rowcoloring}
E10 & Local Entrepreneur & Clothing brand & None & N/A & N/A \\
P11 & Community Provider & \raggedright Educator, \cf leadership & >5 years & Media & ChatGPT, Midjourney \\
\rowcolor{rowcoloring}
E12 & Local Entrepreneur & Marketing & None & \raggedright Website copy, captions, hashtags & ChatGPT \\
P13 & Community Provider & Community Coordinator, \cf staff & 3 months & \raggedright Copy, short-form video content & Capcut, Copy.ai, Speech-to-text \\
P14 & Community Provider & Tech support volunteer & 1 year & Idea and outline generation & ChatGPT \\
P15 & Community Provider & Entrepreneurial support, \cf staff & 6 months & \raggedright Flyer, copy writing, research & ChatGPT \\
\bottomrule
\end{tabular}
\label{tab:demo}
\end{table*}

\section{Findings}
Three core findings emerged from our analysis. 
First, while generative AI technologies are presented with a veneer of user-friendly simplicity, we found the entrepreneurs needed to navigate a large set of operational and strategic skills, as well as engage in ``pre-'' and ``post-processing'' to derive actual and sustained value from these technologies. 
Second, we found that centering communal experience (and supporting non-use) was critical when onboarding entrepreneurs to generative AI technologies in order to overcome overwhelm and techno-anxieties these tools elicit. 
Third, after introduction, we found how local entrepreneurs used generative AI technologies, which technologies they used, as well as initial concerns entrepreneurs had when using generative AI technologies for their business.

\subsection{Taking Stock of Skills Needed for Successful Application of Generative AI}
Throughout the co-design process, workshops, and interviews, providers and entrepreneurs repeatedly pointed to a tension: the way the tools had been marketed and the rhetoric surrounding these tools as simple-to-use magic conveyed an unrealistic expectation that the tools would provide immediate value to users out-of-the-box. 
For instance, after participating in a workshop, E10 reflected on what he observed to be an oversimplification of generative AI online: \textit{``People on the internet kind of simplify AI [saying] if I use AI it will decrease project time lickity split''.} 
\change{As P7 shared, this oversimplification was connected to technologists' assumptions that software tools are ready for use immediately:} \textit{``We do too much of, `Hey here's the software, start using it.' ''}
\change{P7 went on to describe, as a leader at \cf and a tech educator at a local university,} how he frequently witnesses a pattern of inaccurate expectations technologists have for end-users, as they assume their technology is designed well enough for easy use by all. 
While this issue was more generalizable than newly developed AI technologies, participants noted the heightened \textit{``smoke and mirrors''} \change{for how generative AI was portrayed and marketed,} like P4 who shared: \textit{``a lot of these tools are still in beta...there is a lot of smoke and mirrors.''}
Taken together, \change{ participants expressed how the workshops simultaneously needed to make generative AI technologies more approachable by showing entrepreneurs that \textit{``you can do this real quick on your phone''} (P2), while also clearly conveying the amount of work required for actual utility. }
P4 went on to say that, in order to make generative AI tools actually useful for entrepreneurs, \textit{``there is so much wraparound support that is needed.'' }
\change{This \textit{``wraparound support''} is what we detail in the next section.}

\subsubsection{Detailing the ``Pre-'' and ``Post-Processing'' Required}
Part of this \textit{``wraparound support''} included being upfront with entrepre-neurs about the necessary steps required so that these technologies may provide business utility, as P4, an audio engineer and \cf staff shared: \textit{``If you're using artificial intelligence for your accounting, you can't just throw a bunch of receipts in [ChatGPT]. You gotta organize [the receipts] in certain ways. It's the same thing with audio and video. It's the same thing with everything.'' }
This \textit{``same thing''} P4 highlighted was the need for what he later referred to as \textit{``pre- and post- processing''} or the steps that a user must take to organize and clean any inputted data to generative AI tools (pre-processing), as well as the steps required to take any generated outputs and turn them into useful artifacts (post-processing).
In the context of audio and video, P4 detailed how he worked with an entrepreneur to use Vidyo.ai (a generative AI platform for video editing and content creation), which required file type conversations; he listed five different file conversions required to use just one of the entrepreneur's videos recorded from an iPhone. 

Similarly, E3, who provided the demonstration for how he used generative AI to create custom designs for his clothing line, shared that he hoped entrepreneurs would remember the importance of \textit{``getting photos out of DALL-E and into Photoshop, Illustrator or Procreate''} in order to edit the images for them become usable. 
E3 shared that \textit{``nothing is ever picture perfect off the Internet,''} and he noted that the images generated were helpful for a brainstorming step.
However, in order to use the generated images for his business, E3 detailed the variety of tools needed to edit the images and change file types before they could be sent to his clothing supplier (e.g. he used Photoshop to edit and overlay images onto apparel mockups). 
\change{Entrepreneurs in the audience during E3's demonstration took note of these steps.
For instance, one entrepreneur shared how this perspective of generative AI technologies was both inspiring and sobering,} specifically noting E3's time investment: \textit{``just seeing how [E3] was able to create his own clothing line. And how much he had to [do], basically going through a ton of different images...who knows how long it took him?'' }
Taken together, P1, P2, P4, P7, and E3 all highlighted the importance of the workshops to make salient the full work that was required to use these tools effectively; only then would entrepreneurs be able to cut through the often disillusioned marketing and rhetoric surrounding generative AI and decide for themselves if these tools would work for them and their business. 

\subsubsection{The Steps which Preempt Prompt Libraries}
Of course, when it came to introducing generative AI technologies to local entrepre-neurs, reviewing prompt engineering techniques was central. 
As described in Section \ref{sec:workshopdesign}, we leveraged state-of-the-art prompting techniques such as personas~\cite{white2023prompt}, and populated example prompts related to each workshop theme.
However, we found that providing prompt libraries, both paper and digital copies, was typically ineffective during introductory workshops. 
As P15 shared, \textit{``we were skipping steps with having the prompts on paper, prompt [engineering] in itself is its own workshop. It's its own tool that people have to learn how to use. What I found, especially working with [E9], it was easier to architect a prompt as I was explaining to her what a prompt is.''} 
Here, P15 described their interaction with E9, whom they were partnered with during a workshop.
Together, they co-articulated the prompt as they discussed the nature of a prompt.
P15 went on to explain how \textit{``without [entrepreneurs] having some sort of frame of reference in general...how can you be a prompt architect if you don't know the nature of AI? How do you tell it to give you a 4k image without people inside if you don't know what a prompt is?''} 
Similarly, E9, E12, P14, and P15 discussed how it was critical for entrepreneurs and providers to engage in back and forth discussion about entrepreneurs' businesses, business goals, as well as their technology goals, and then write initial prompts together to \textit{``see tangible outcomes for what AI can do for them, so that they know what it is.'' }
\change{As described in the next section, breaking down prompt engineering into the distinct operational skills required was essential to ensure workshops provided adequate support, as well as supporting the other skills beyond prompt engineering needed for successful use.}

\subsubsection{A Laundry List of Operational Skills Needed to Use Generative AI Technologies}
P1 dissected the operational skills required for prompt engineering: using language in a dialogue box, entering a search query (or prompt), and then knowing how to iteratively refine a search query (or prompt) to get better results. 
However, beyond prompt engineering, P1, P2, E3, P4, P7, and P11 compiled a list of foundation, digital skills entrepreneurs should acquire for successful use of generative AI technologies: basic browser literacy such as opening a new tab and switching between tabs and windows (such as when moving between tools for pre- and post-processing); understanding file types and file conversions in order to transition files between generating, editing and publishing software; understanding files systems both locally and cloud-based to save generated results; understanding storage management to edit the output or share with providers, employees, customers, mentors, or peers; word processing system skills in order to customize generated text; graphic design knowledge to take generated images and \textit{``put it into Canva''} or other visual editing software to edit photos effectively (i.e., online aesthetics and design guidelines); password management skills to sign in and out of various tools; typing skills and keyboard shortcuts to quickly enter and edit prompts, as well as copy and paste generated media. 

Taken together, this list included seven operational skills that entrepreneurs needed in order to effectively use generative AI tools for their business. 
Beyond these operational skills, P1, P2, P4, P7, P11, and P13 expressed the importance of entrepreneurs' sense of self-efficacy when using generative AI. 
For instance, when discussing the workshop format, P1 shared, \textit{``Start with self-efficacy. `The tool is not above you. It's not smarter than you. You have the intelligence inherent within you. You have the wherewithal to leverage these tools just as much as anyone else.'''} \change{Here, P1 spoke to the importance of emphasizing entrepreneurs' self-efficacy as he described how workshops can directly mitigate any doubts entrepreneurs may have when it comes to their abilities to use a novel technology.
Similarly to P1, P13 further emphasized the importance of self-efficacy messaging throughout the workshop series as he shared:}\textit{``That's all [entrepreneurs] need, their brain...Everyone has the capacity within them to learn this skill.''}
\change{By taking into account self-efficacy and often overlooked skills, the workshop series made strides towards achieving community-driven objectives, as described in the next section.}

\subsubsection{The Larger Goal: Supporting Use, and Non-Use} 
Providers (P1, P2, P4, P7, P13, P14, P15) emphasized that workshops needed to support a range of levels of use of generative AI technologies: supporting entrepreneurs who wanted to dive in and incorporate generative AI throughout their business processes, as well as supporting entrepreneurs who were interested to try out generative AI technologies, but ultimately may decide not to use them. 
For instance, P13 emphasized how it was important that the workshops were \textit{``not putting [generative AI] in the participants' face[s]''}, \change{and instead workshops provided various other activities and peer networking opportunities (See Table~\ref{tab:workshopoverview}).}
Instead, P15 viewed the ideal outcome of the workshops to be \textit{``a frame of reference,''} for entrepreneurs to have a sense as to what generative AI technologies are, and what these technologies can do for their business. 
Similarly, P1, P2, P4, and E10 emphasized that exposure to the technology was more important than being concerned with immediate proficiency, as a matter of entrepreneurial agency: \textit{``If you know it exists, [then] you can decide how to use it.''}
P4 continued on to share that workshops needed to \textit{``show [entrepreneurs] what the AI is capable of [and] not capable of, show all of the pre- and post- processing. Don't blow smoke. It might make [entrepreneurs' tasks] better, faster, [or] it might take longer.'' }
\change{As P4 described, a sobering perspective of generative AI technologies was essential in order to align workshops with community-driven goals to support both use and non-use. 
In either case, as described in the next section, centering entrepreneurs' shared experiences and commonalities such as geographic proximity or affiliation with \cf was an important way to structure introductory workshops.}

\subsection{The Importance of Supportive Exposure as a Communal Experience}
Community providers and local entrepreneurs frequently discussed the importance of communal and supportive exposure to generative AI technologies in the context of entrepreneurship (P1, E3, E5, E6, P7, E8, E9, P13, P14, P15).
Since entrepreneurs were from the surrounding community, and the workshops took place in a shared community space, this provided commonality for entrepreneurs to build a shared experience.
P1 argued this was essential to successfully introduce entrepreneurs to generative AI: \textit{``all the people attending are members in the local society, the local economy, all of them are entrepreneurs...They also have familiarity with \cfns. [The workshops] are building commonality right there.'' }
\change{As P1 described, building commonality through a shared community space was important for both during and after the workshops; in several cases, after connecting during workshops, entrepreneurs continued to meet at \cf or elsewhere locally to provide ongoing support or even collaborate.}

\subsubsection{Overcoming Overwhelm, Together}
\change{Centering communal experience during workshops (through a shared space, round robin introductions, ice breakers, peer networking, small group work sessions, and so on) was important for several reasons.}
First, communal exposure helped to provide a safe space to navigate a range of emotional responses participants experienced towards generative AI technologies. 
Entrepreneurs expressed anxiety about being left behind by not understanding or using the technology (E3, E5, E6, E8, E9), sharing \textit{``whether I like it or not, technology is the future. You have to be equipped or you are going to be left behind''} (E5), and \textit{``jump on the train before [you're] left behind''}(E8). 
Entrepreneurs also expressed being intimidated and fearful of generative AI technologies. For instance, E6 reflected on the first time she opened ChatGPT, and, upon seeing the text response, immediately closed the window and shut her laptop: \textit{``someone told me about ChatGPT a while ago...When [I opened] it up, and then I put something in there, and it just spit out all this information. I was overwhelmed. I was like, `Nope, it's not for me. Log off.''' }
Similarly to E6, before the workshop, E9 shared that her impression of generative AI technologies was primarily fear-based: \textit{``It was total fear...the fear of the unknown kept me from jumping in sooner.''}

As with E6 and E9, E8 shared how the first workshop he participated in was primarily memorable for the experience of seeing others in a similar situation; this helped him to overcome a sense of overwhelm he felt towards generative AI technologies: \textit{``I won't lie, the first session was a little overwhelming. But it was just nice being in the space where everybody was all coming from the same place, just trying to learn and get an understanding of what we have in front of us.''}
\change{In this way, the first workshop for E8 was impactful because he witnessed that he was not alone by observing others' trepidation and excitement within a shared space.}
He went on to share: \textit{``The second session was more impactful because I had time to absorb what [generative AI] can do, things of that nature.''} 
\change{As E8 described, after the first workshop where he established a sense of community, he was better able to digest information in the following workshops.}
E8 shared how, one of the critical aspects of both workshops he attended was how the workshops provided a ``safe space'': \textit{``It was a safe space that everyone is...unfamiliar with, and they just wanna get familiar [with AI tools]. While it is overwhelming...it doesn't linger `cause then you can look around the room and see people just like you in the same space...there's no stupid questions.'' }
\change{Here, E8's reflection makes salient a few aspects of the workshops which contributed to a sense of safety.
For instance, E8 felt comfortable to ask questions freely and without judgement from peers or providers.
In addition, he witnessed other local entrepreneurs' reactions and questions as they digested information about the technologies for the first time, too.
Taken together, E6, E8, and E9's reflections highlight the importance of convening in a community space in order to overcome the overwhelm associated with generative AI technologies, together.
}

\subsubsection{Building Long-Term Community and Technical Capital}
\change{In addition to overcoming overwhelm, communal experience helped to catalyze technical capital building and continued support among entrepreneurs and \cfns.
This was especially important when onboarding entrepreneurs who were less likely to be embedded in a network of ``techies'', as P7 described:} \textit{``if you're not a techie, I think it really helps that you have a community to explore this craft and tool with...especially if you're in a community that's generally not using [AI] tools.''} 
\change{Here, P7 pointed to the importance of being embedded in a technical network, where members can easily share and discuss new software releases, hacks, problems, and more, providing both critical information and encouragement.}
One way that the workshop series supported technical capital and long-term support was by connecting the workshop structure to other ongoing technical services at \cfns.
For instance, workshop attendees were encouraged to partake in the weekly (free) technical office hours, called \thdns, hosted at \cf for continued support between and after workshops; several entrepreneurs from the workshops (including E5, E6, and E9) attended multiple offerings for continued support. 
P13 reflected on the role of this continued support: \textit{``\thd is a piece of [the workshops] that is super important. It allows entrepreneurs a way to come back and stay engaged [and is] really impactful in terms of meeting people where they are.''}
\change{With this continued engagement, P14 observed that by the fourth workshop,} \textit{``Most of the people involved were not strangers to [each other]...it's important to do something like this in a community that they feel safe with.'' }
\change{For P1, building technical capital through a communal onboarding experience played a role in claiming power, especially among entrepreneurs who may have been systemically disempowered and under-supported to use novel technologies: }\textit{``Where do [entrepreneurs] go to get instruction on how to claim power? We have to go to our community, and strategize within our community...you have to be around people who have skill and strategy of how to be effective with the tool.''}
\change{Importantly, P1 emphasized the role of within-community expertise to embolden entrepreneurs who historically have been disempowered by novel technologies.}
One way the workshops built on this point and showcased \textit{``skill and strategy from within the community''}, was through a live demonstration from a local entrepreneur who was also a \cf member, as detailed in the next section.

\subsubsection{Showcasing In-House Expertise}
\label{sec:inhouse2}
As described in Section~\ref{sec:workshopgoals}, the fourth workshop featured a local entrepreneur's (E3) live demonstration of how he used generative AI for his business.
Almost all participants who attended this workshop commented on its impact, especially because E3 was a member of \cfns.
For instance, P7 commented: \textit{``the presentation by [E3] was really powerful...for people to see a young African-American dude who [is] around \cf as an [entrepreneur], here's how quickly he made a product that he's selling using an AI tool...I even audibly heard people being like, `whoa.''' }
\change{Here, P7, who was in the audience during E3's demonstration,  commented on the auditory reactions audience members had while witnessing the live demonstration of E3 using ChatGPT to generate complementary color schemes via HEX codes, DALL-E 2 to create images of the Pittsburgh skyline, and then Procreate and Photoshop to edit images and overlay creations onto clothing mockups.}
E9, who had a custom gift basket and event planning business, shared how E3's demo inspired her and helped her better understand what was possible with tools like DALL-E 2 and ChatGPT: \textit{``To be able to hear how that young man used AI to create the shorts, it had my mind spinning. Now that I know you can do that, I want to be able to make my own ribbon, a custom banner.'' }
\change{Taken together, showcasing in-house expertise presented an opportunity to further bolster technical capital building and create an empowered sense of community which was critical to overcome overwhelm associated with generative AI technologies.
From this standpoint, entrepreneurs could then more confidently decide how to use (or not use) these technologies for their business, as detailed in the next section.}

\subsection{Entrepreneurs' Initial Uses (and Non-Use) of Generative AI, and Concerns}
Participants shared their initial uses of generative AI technologies to make logos and flyers, write grant applications, optimize search engine results, and more (See Table~\ref{tab:demo}).
In addition, entrepreneurs shared how their preferences developed with respect to how they did not want to use  generative AI for their business, such as in the case of handling material which was sensitive (e.g., concerns relating to sentimentality, intellectual property, inauthenticity, and biases). 

\subsubsection{How Entrepreneurs Used Generative AI for Business}
Entre-\\preneurs used ChatGPT in a variety of ways to boost their business' marketing capabilities upon taking a workshop.
E5 referred to ChatGPT as a \textit{``24-hour secretary''}, and he described asking ChatGPT to become a content creator for his social media accounts. 
E5 estimated that he used ChatGPT almost daily, and was still experimenting and getting comfortable formulating persona-based prompts such as, \textit{``You are a fortune 500 CEO...'' ``you are an owner of a mom and pop store...'' }
E8 used Canva Text-to-Image and DALL-E 2 to make a logo for his podcast, which used his initials: \textit{``I don't consider myself the greatest creator. I wouldn't say I'm very imaginative or anything like that. So that's where the software helped me...I just put in [my initials]...and I just wanted to see what different variations it would create for me. And it did!'' }
During a workshop, E12 used DALL-E 2 to create a flyer for an upcoming business event.
Specifically, she made an image of people drinking wine on a roof during sunset, where she started with a basic rooftop and then included ``tech'' and ``future'' in the prompt to make the resulting image more futuristic. 
During a workshop, P2 showed an entrepreneur with a hair braiding business how to generate images of African-American women with different kinds of hairstyles, and then how to add her logo to the images, all on her mobile phone with Wonder.ai and Photoroom. 
Between workshops, E6 attended multiple offerings of \thdns. 
After working on a social media marketing plan during a workshop, she sought continued guidance on writing an application for a local small business grant using ChatGPT. 
When asked what her vision was for how to incorporate AI in her business, she noted that in the immediate future, she wanted to use ChatGPT to improve her website's SEO and lead generation (e.g. through blog writing), and create designs for flyers. 
E9, another consistent attendee of \thd and the workshops, used the Wix.com AI-generated product descriptions to zhuzh her existing product descriptions within the website editor interface: \textit{``I used it to reword my descriptions...For gift basket descriptions, if you want to reach a broader audience, your wording needs to cover a wide variety of people.'' }
For E9, Wix's generative AI feature helped her to add more detail that she had not considered such as other types of customers who may be interested in a gift basket. 

\subsubsection{\change{Reasons for Non-Use:} Participants' Concerns for Using Generative AI for Business}
\label{sec:concerns}
Entrepreneurs expressed several concerns when it came to using generative AI for their business, \change{and sometimes these concerns led them to opt not to use these technologies.} 
Detailing such concerns is critical, in order to be able to develop a curriculum which can formally and proactively address concerns by equipping entrepreneurs with relevant skills to navigate these concerns, and support critical non-use. 
One common concern entrepreneurs expressed was whether the use of generative AI for content creation would make it harder to convey their business' unique brands (E3, E5, E6, E9, E12). 
For instance, E6 emphasized that she wanted to \textit{``make sure that I'm authentic.''} 
To achieve this, she actively edited the synthetic text to customize it and to make sure it was accurate, and also used a specific persona in ChatGPT: \textit{``I put the role in the text, but I'm very specific. I go back and forth about how authentic it is...Let me go back and rewrite this, personalize it, look [it] up to make sure these numbers are accurate.''}
E9 echoed similar authenticity concerns to E6, \change{however, her concerns amounted to her decision to not use tools like ChatGPT to communicate with her customer base:} \textit{``I never want people to read my own stuff and say, `oh wow, she doesn't even write her own stuff.' ''} 
\change{For E9, her underlying concern was that she would lose her unique communication style with her audience and ultimately damage her reputation.}
\change{Taken together, E6 and E9's shared concerns reflected two issues with generated outputs.
First---as an issue of the quality of generated outputs---E6 needed to repeatedly revise her prompts such that the generated outputs were to her standards and this took considerable effort.
Second, even with revised prompts, E9 expressed a fundamental concern that her audience may perceive text as synthetic and judge accordingly, and decided to remove this risk entirely by not using the technology.}
E12 shared an internal dilemma as well, as she reflected on when it was and was not ``appropriate'' to use generative AI, which was partly informed by whether she thought the artifacts needed to be created by her, or not. 
In particular, she focused on whether AI-generated content could be construed as false advertising, and if so, she wanted to avoid use.

Entrepreneurs and providers also expressed anticipated concerns for long-term use such as becoming overly dependent on generative AI (P1, E3, E9, E10, P11, E12).
For instance, E10 shared that, over the long term, he was \textit{``afraid that people might see AI less of a tool and more of a co-dependent.'' }
If co-dependence prevailed, P1 noted that long-term use may contribute to feelings of imposter syndrome, where entrepreneurs may attribute any business success to generative artificial intelligence over their own, triggering thoughts such as: \textit{``Did I cheat? Do I have something to hide?''}
E10, one of the participants in our study who reported he was less likely to use generative AI for his business shared: \textit{``I do not want to become brain dead in the future. You still have to know the general basis of your audience.'' }
P1 and P11 detailed how this technology was similar to other technologies like auto-fill and predictive typing, where now, users \textit{``don't need to know how to type...This is an example of tool that came to help us, but in its use we have lost the ability to [spell].''} 
To address this concern, P11, leadership at \cf who developed tech curriculum for youth, emphasized the importance of maintaining balance between tool use and skill development: \textit{``We try to be balanced. We recognize there is an opportunity [and] teach the complementary skills that AI is taking away from [users].''}
In doing so, P11 went on to say that use of generative AI needs to be reframed from \textit{``outsourcing''} to \textit{``collaboration.''}

\change{In addition to perceived inauthenticity, accusations of false advertising, and overreliance,} participants detailed additional concerns \change{which motivated non-use} focused on ethical and legal dilemmas such as issues of ownership and intellectual property (E3), transparency of use (P1), racial and gender bias (P1, P2, P14, P15), and tech and data extraction (P1, P4, P14, P15). 
For instance, E3, who demonstrated how he used DALL-E 2 and ChatGPT to create imagery for his clothing line, considered the role of ownership: \textit{``I believe OpenAI has it so that you actually own the images that you generate with your prompts, so legally they're your property now. So I feel like in that area [it] is fine.''}
Based on OpenAI's current policy, users do own the images they create, regardless if you used free credits or paid~\cite{sell}.
However, during the process of writing this paper, a federal court ruling prohibited copyright of AI generated art~\cite{court}, which speaks to the rapid evolution of policy regulating this novel technology~\cite{jiang2023ai}, and the need to keep entrepreneurs' up-to-date on legal considerations of use. 
Even with this in mind, E3 had additional benchmarks for ethical use: \textit{``I feel like as long as I'm not taking the image, putting it on like a T-shirt, and then screen printing that T-shirt, doing the bare minimum [and] gaining profit from it, then I feel like we should be okay since nobody was hurt in the process.'' } 
He continued on to state how this benchmark was based on a version of the ``golden rule'': he did not want someone else to query: ~\textit{``create art in the style of [E3]''} and then be able to sell or own the generated images, so he applied the same principle to himself. 
For E5, who was authoring a book based on his experiences while incarcerated, he contemplated using ChatGPT to help with copy editing and proofreading, but decided against it because the topic was too personal.
It was unclear to him how his experiences and emotions would be used or extracted by the technology. 

In addition to legal concerns, P14 reflected on how these generative AI technologies perpetuate systemic biases based on the biases embedded in training data. 
In particular, she reflected on one instance when an entrepreneur in the workshop uploaded a photo of their late relative to inspire a logo design. 
This reference image was a portrait of a Black woman who had short black curly hair, yet the resulting images generated by Midjourney included a woman with light, straight hair pulled back into a loose ponytail. 
While the entrepreneur and provider were able to work together to create a prompt which did not white-wash the reference image, this experience highlighted critical issues with these tools.
P14 continued:  \textit{``I am very conflicted around these tools, how they collect data, privacy concerns, how big tech is so extractive. But that all goes out the window, if you want community input, they need to use these tools...gatekeeping information is actually really detrimental.'' } 
P13, who worked with the same entrepreneur after the workshop on her business's logo, shared that the entrepreneur decided to work with a graphic designer instead due to the sentimental nature of the photo and the need for a careful human touch. 
\change{Taken together, entrepreneurs and providers shared several concerns for using generative AI for their business.
And, as illustrated in the above examples, sometimes these concerns amounted to entrepreneurs deciding to not use these technologies at all.}

\section{Limitations}
There were several limitations of this study. 
\change{First, while recruitment targeted all workshop attendees, we observed that the entrepreneurs who opted to participate in interviews tended to be more engaged in the workshops; this biased data collection towards positive experiences.
We attempted to offset this bias in two ways.
First, we solicited critical feedback from community providers with whom we had long-standing rapport.
Second, we detailed many examples where entrepreneurs who, overall, were open to using generative AI for their business, yet decided certain use cases were not appropriate. 
Relatedly, a second challenge was that not all entrepreneurs wanted to engage with generative AI technologies. 
While we frame this lack of use as less of a limitation, and instead an important consideration~\cite{baumer2014refusing}, we understand that this reduced the quantity of usage data collected.}
\change{As a result, our study presented limited data on the quality of generated outputs.
Future work can analyze entrepreneurs' assessments of quality of generated outputs, accounting for novelty effects by analyzing use over time}.
Finally, this study was exploratory in nature, and we did not formally assess knowledge transfer. 
Instead, we provide initial insights that could be developed in future work to assess the efficacy of a more formalized curriculum for generative AI in the context of entrepreneurship. 


\section{Discussion}
This work detailed the importance of social support when onboarding local entrepreneurs to generative AI. 
Through a close partnership with \cfns, we co-designed an interactive workshop series focused on meeting entrepreneurs where they are, provided actionable and tangible outcomes, and embedded technical support in a network of vetted and trusted relationships. 
Our approach enabled us to gain knowledge about (RQ1) the importance of centering communal experience when introducing generative AI to support local entrepreneurs’ due to a range of comfort levels, (RQ2) the importance of cutting through hyped rhetoric to provide entrepreneurs with a practical understanding of the work (and time and skills) required to derive value when applying generative AI technologies for business, and (RQ3) an early look at how local entrepreneurs use---and prefer not to use---generative AI for their business.

\change{In answering these research questions, this paper makes three core contributions. 
First, by building on models of low-tech social support~\cite{dillahunt2018entrepreneurship, hui2018making, avle2019additional, raval2019making, pei2020attenuated, hui2020community, ogbonnaya2019towards, kotturi2022tech}, we presented empirical findings for an interactive workshop series tailored for generative AI technologies.
With an eye towards future work, in this section we consider other forms of low-tech social support---beyond small, local networks---which may uniquely respond to the need to keep pace with rapid technological advancements and policy changes related to generative AI technologies.
Second, we contribute empirical findings of local entrepreneurs' use of generative AI technologies, as well as their concerns as usage relates to their business and motivations for non-use.
In this discussion section, we further consider how future work can create more formalized AI literacy curricula tailored for local entrepreneurship as well as more formally support concerns and instances of non-use.
Finally, we contribute the details of an approach to designing community-driven workshops~\cite{cooper2022systematic} such as building on a four year and ongoing tech support program, embedding the workshop series in ongoing community initiatives, and ensuring value generation for the community center by emphasizing community-driven data collection and goals.
We argue that our approach was essential to conducting this study given the techno-anxieties---nervousness and apprehension~\cite{marcoulides1989measuring}---entrepreneurs shared.
Taken together, in this discussion section, we discuss these themes and contributions to formulate considerations for future work, interweaving feedback from entrepreneurs and providers on suggestions to improve the workshop series. }

\subsection{Deconstructing the Veneer of Simplicity}
Community providers and entrepreneurs in our study surfaced a tension between the way pre-trained generative AI was marketed to entrepreneurs by platforms and self-proclaimed experts alike, versus the \textit{actual} time, skills, and even additional tools needed to move beyond superficial interaction towards valuable use. 
We consider this tension alongside current disparities of use where generative AI technologies like ChatGPT and DALL-E 2 are primarily used by those who have backgrounds in technology or are college educated~\cite{pew-aug28}.
In particular, this tension merits inspection of the kinds of digital skills required to leverage such innovative technologies.

Digital literacy involves various types of skills such as strategic skills---high-level, goal setting---and operational skills---low-level implementation~\cite{van2011internet,ala2011mapping}.
Given rapid technological advancement and recent end-user access, digital literacy in the context of generative AI has not yet been clearly defined, inside nor outside of a classroom context~\cite{ng2021conceptualizing}.
For instance, some say that the operational and strategic skills involved with using generative AI technologies effectively are similar to those required for effective online search~\cite{metzler2021rethinking}, involving periods of exploration and exploitation as in an information retrieval task~\cite{pirolli1999information}.
However, other researchers argue that this analogy is flawed given the differences between information retrieval and information generation. 
For instance, Shah and Bender investigated how chat-based interfaces limit a user's ability to engage in typical sensemaking which is afforded in online search (e.g., comparing across other links, identifying patterns to detect misinformation)~\cite{shah2022situating}.
They argued that because generative models often lack references, predominant interface design detracts users’ abilities to verify information generated.
To start to articulate general framework for AI literacy, Ng~\etal conducted a literature review of AI literacy in education to create a framework highlighting four levels of competency: ``know and understand AI, use and apply AI, evaluate and create AI, ethical implications of AI''~\cite{ng2021conceptualizing}.
In this framework, it is assumed that these are levels which must be acquired linearly for competency. 
However, their work focuses on a pedagogical context and is, therefore, only partially applicable in the context of entrepreneurship in lean economies. 
For instance, in our study, entrepreneurs wanted to ``know and understand AI'' in the context of entrepreneurship (not necessarily focusing on the details of how diffusion models work), and shortly after this level, entrepreneurs considered the fourth ``level'': ``ethical implications of AI''. 

\change{When considering a framework for AI literacy in the context of local entrepreneurship,} this study provided a preliminary and exploratory opportunity to surface the skills needed which ``democratizing'' promises of tech usually overlook~\cite{hui2020community,von2006democratizing}.
In doing so, we found that there were many more skills than simply prompt engineering involved (and during initial use, prompt libraries were less helpful). 
The additional strategic and operational skills entrepreneurs needed to have to use these tools---on top of access to Internet-enabled devices---included: browser literacy, understanding file types and file conversions, storage management literacy, word processing skills, and more.
The workshop series explored one way to support these skills through a communal experience (e.g. providing devices, pre-logging into platforms, co-articulating prompts).  
Here, we are not suggesting a finalized curriculum, but instead are sharing insights that could be used to support more formalized curriculum development focused on equity in future work. 
\change{In this way, future work can develop a more formalized approach to supporting AI literacy in the context of local entrepreneurship.
For instance, building on the importance of meeting entrepreneurs where they are~\cite{gautam2020crafting}, researchers can create AI literacy modules embedded in digital and physical community spaces such as community centers' websites and in-person workshops which target specific literacies required for successful use of AI technologies.}

In addition to providing an initial look at the skills needed to effectively use AI tools, it was also critical that the workshops supported self-efficacy of entrepreneurs and demystified generative AI technologies through collective inquiry. 
After entrepreneurs’ established a ``frame of reference'', only then could discussions of larger implications of using generative AI for business occur. 
We related this premise to the ``Consentful Tech Project''~\cite{consentful}, which focuses on tech use which is freely given, reversible, informed, enthusiastic and specific (FRIES), building on the feminist idea of enthusiastic sexual consent~\cite{strengers2021can}.
In our study, many of the providers were privy to the many ethical implications of generative AI and discussed the importance of first having entrepreneurs, who were less familiar with generative AI technologies, to be introduced within a communal space. 
Having a clear understanding of the issues as a form of consent, we reflect on the concerns entrepreneurs had for using generative AI technologies soon after introduction, as detailed in follow-up interviews. 
For instance, entrepreneurs expressed concern for becoming overly reliant and dependent, damaging their reputation and becoming disconnected from their audience, being subject to intellectual property infringement, as well as concerns due the systemic bias embedded in generative AI technologies. 
While there have been studies which find end-users are overly reliant on these models~\cite{wang2023decodingtrust}, we found that entrepreneurs in our study were actively, and intuitively, skeptical of these technologies even if they were unsure of how to change their use to address their concerns.
Future work can investigate how to support entrepreneurs’ critical use of generative AI technologies.
\change{In particular, future work could explore how to more readily illuminate entrepreneurs' reasons for non-use or limited use as a way to highlight key concerns, and provide strategies for entrepreneurs facing similar dilemmas.
For instance, building on HCI scholarship which focuses on digital storytelling~\cite{halperin2023probing}, researchers could collaboratively create a repository of anonymous stories of how entrepreneurs choose to not use generative AI technologies for their business, as well as the reasoning behind this decision. 
Such a repository could provide nuanced and granular documentation of the everyday tensions entrepreneurs navigate with generative AI technologies as a form of distributed and scalable mentorship~\cite{hui2019distributed}.}


\subsection{Extending the Role Social Support When Onboarding Local Entrepreneurs to Generative AI}
Both entrepreneurs and community providers reflected on the opportunity for generative AI to level the playing field in entrepreneurship, but many were simultaneously skeptical. 
\change{Such skepticism was well-founded and reflected the often superficial and unwarranted promises technologists make of more democratic futures with each technological advancement~\cite{hui2020community}.
As our providers and entrepreneurs described, \textit{actual} equitable tool use required wraparound support, such as continued engagement with \thdns, low provider-entrepreneur ratios, and building a sense of community through shared experience.
Here, we first revisit Hui~\etal’s recent model of low-tech social support for maintaining digital engagement for entrepreneurs~\cite{hui2020community},
and consider how our findings build on this model.
For instance, similarly to Hui~\etal's model, the workshops benefited from small scale (to encourage trust building), resource connecting organizations (such as with \bsc and \thdns), paper planning tools (which entrepreneurs were provided with if they did not bring them), regular in-person meetings (both the workshops and \thdns), and validation and practice.
}

\change{However, our findings also provided an initial understanding of the unique aspects of wraparound support that were critical when onboarding entrepreneurs to generative AI technologies in particular, such as parsing instances when generative AI may be more or less appropriate to use given fast-paced developments and rapidly evolving policy~\cite{court}.
In these cases, we consider how small, local networks may find it challenging to keep pace with these rapid advancements (as did our university-community team), and therefore we suggest that future work explore community-driven technology interventions to pool unfolding information (such as a policy changes) and extract relevant implications to be shared with local, small business owners about these advancements.
To incorporate community perspectives, future work could leverage Erete’s framework for designing community-driven technological interventions~\cite{erete2014community}, and apply intersectional analyses of power in design~\cite{erete2023method}.
In particular, such scholarship provides a road map for designing technology which acknowledges the structural oppression and institutionalized racism embedded in HCI and computing more generally.
}

\section{Conclusion}

This paper investigated how to onboard local entrepreneurs to generative AI technologies through a community-driven protocol which built on a four-year relationship between university and community team members.
In doing so, we highlighted the importance of centering communal experience when introducing generative AI to support local entrepreneurs’ range of comfort levels and situate technologies in shared experience.
In addition, we detailed the importance of cutting through hyped rhetoric to provide entrepreneurs with a practical understanding of the actual work, time and skills that were required for successful application. 
Taken together, this paper provided an early look at how entrepreneurs used generative AI, and how they preferred to not use these technologies for their business.
And while we are not the first researchers in HCI to describe how technology is embedded in and intertwined within entrepreneurial practices, this age of rapid apparent technological change may cause some to question whether this time is different. 
Our answer is a resounding ``no’’, however, the details and particulars are different in important and impactful ways, as this research begins to unpack.

\begin{acks}
This work was funded in part by the National Science Foundation (1928474). We thank the entrepreneurs and providers who were involved in the workshop series, and all of the \cf staff and leadership for their support, especially TJ Johnson, Amil Cook, Patrick Cooper, Arturo Lozano, Hamza Perez, Bethany Hallam, and Nora Reed.
\end{acks}

\bibliographystyle{ACM-Reference-Format}
\bibliography{00-main}


\end{document}